\documentclass[fleqn,usenatbib]{mnras}

\usepackage{newtxtext,newtxmath}

\usepackage[T1]{fontenc}

\usepackage{listings}

\DeclareRobustCommand{\VAN}[3]{#2}
\let\VANthebibliography\thebibliography
\def\thebibliography{\DeclareRobustCommand{\VAN}[3]{##3}\VANthebibliography}

\usepackage{graphicx}	
\usepackage{amsmath}	

\title[JAXNS: high-performance nested sampling in JAX]{JAXNS: a high-performance nested sampling package based on JAX}

\author[J. G. Albert]{
Joshua G. Albert,$^{1}$\thanks{E-mail: albert@strw.leidenuniv.nl}
\\
$^{1}$Leiden Observatory, Leiden University, PO Box 9513, 2300 Leiden, The Netherlands
}

\date{Accepted XXX. Received YYY; in original form ZZZ}

\pubyear{2021}

\begin{document}
\label{firstpage}
\pagerange{\pageref{firstpage}--\pageref{lastpage}}
\maketitle

\begin{abstract}
Since its debut by John Skilling in 2004, nested sampling has proven a valuable tool to the scientist, providing hypothesis evidence calculations and parameter inference for complicated posterior distributions, particularly in the field of astronomy.
Due to its computational complexity and long-running nature, in the past, nested sampling has been reserved for offline-type Bayesian inference, leaving tools such as variational inference and MCMC for online-type, time-constrained, Bayesian computations.
These tools do not easily handle complicated multi-modal posteriors, discrete random variables, and posteriors lacking gradients, nor do they enable practical calculations of the Bayesian evidence.
An opening thus remains for a high-performance out-of-the-box nested sampling package that can close the gap in computational time, and let nested sampling become common place in the data science toolbox.
We present JAX-based nested sampling (JAXNS), a high-performance nested sampling package written in XLA-primitives using JAX, and show that it is several orders of magnitude faster than the currently available nested sampling implementations of PolyChord, MultiNEST, and dynesty, while maintaining the same accuracy of evidence calculation.
The JAXNS package is publically available at \url{https://github.com/joshuaalbert/jaxns}.
\end{abstract}


\section{Introduction}

Nested sampling \citep[NS;][]{2004AIPC..735..395S} is a Bayesian method of computing general calculations of the form,
\begin{align}
    Z = \int_\mathcal{X} \mathcal{L}(x) \,\mathrm{d}p(x),\label{eq:1}
\end{align}
where $p(x)$ is the Lebesgue measure of a random variable, $x \in \mathcal{X}$, and the integral is a Lebesgue integral of some bounded, non-negative, measurable function $\mathcal{L}: \mathcal{X} \to \mathbb{R}^+$.
In the context of this manuscript $\mathcal{L}(x)=p(y|x)$ is the likelihood function of some set of observables, $y$, in terms of a random variable $x$, and $p(x)$ is a prior distribution for $x$.
In this case, the choice of likelihood and prior is said to constitute a model and $Z$ is the Bayesian evidence of the model.
In addition to computing evidences NS also provides weighted samples of the posterior distribution, $p(x\mid y) = p(y\mid x) p(x) / p(y) = Z^{-1} \mathcal{L}(x) p(x)$.
Since its introduction in \citet{2004AIPC..735..395S}, NS has become an important tool in many sub-domains of astronomy including cosmology \citep{2006ApJ...638L..51M, 2007ApJ...660L..81E, 2009ApJ...699..985H, 2020A&A...641A..10P}, compact high-energy astrophysics \citep{2019ApJ...882L..24A, 2020arXiv200408342T}, spectra modelling \citep{2010MNRAS.401.2531A, 2014A&A...564A.125B}, gravitational lensing \citep{2010MNRAS.408.1969V, 2011MNRAS.415.2215B}, and more \citep{2011ApJ...729..106T,  2013ApJ...764..155L}.

Science is primarily concerned with two aspects of data analysis. 
Arguably, the primary aspect is hypothesis comparison, canonically done via comparison of the evidences, and the second aspect is posterior distribution estimation.
As written in \citet{skilling2006}, `Giving the value of $Z$ (the evidence) is a courtesy to other workers who may wish to perform model selection, and ought to be a standard part of rational enquiry.', though this is often neglected due to the difficulty in computing the Bayesian evidence.
With the advent of NS, and publicly available implementations such as MultiNEST \citep{2009MNRAS.398.1601F}, PolyChord \citep{2015MNRAS.453.4384H}, DIAMONDS \citep{2015EPJWC.10106019C}, and dynesty \citep{2020MNRAS.493.3132S}, especially those with pythonic bindings, the barrier to computing the evidence has been significantly lowered.
The second aspect, estimating the posterior distribution, is also a worthy element of science, not to be downplayed, for this, in combination with some aesthetic principles from the scientist \citep{2005AIPC..803....3S}, is generally how we form priors for future models.
Nested sampling also has a key differentiating factor compared with other work-horse inference methods like MCMC, in that NS is able to handle complicated, multi-modal, degenerate posterior distributions with relative robustness \citep{2008MNRAS.384..449F}, even being able to handle notoriously difficult phase change problems where $\log \mathcal{L}$ contains first-order, non-concave, phase transitions \citep{2004AIPC..735..395S}.

Given the apparent power of NS, either for computing the evidence of a model or estimating complicated posterior distributions, it is important to ask in which situations the current implementations of NS are amenable. 
All of the use cases of NS in astronomy listed above require significant compute time.
This time frame can be on the order of hours to weeks for a single evidence calculation, depending on how expensive the particular likelihood evaluations are and how efficient the NS implementation is.
While these types of inference problems are common in data science, there are classes of time-constrained inference problems that are not able to apply NS in any of its current implementations.
These problems require inference to be fast enough to meet some type of time criterion.
Examples include real-time calibration, huge multi-dataset problems, and serving applications with Bayesian inference in real-time.
Currently, these problem classes are not serviced by NS because all NS implementations are too slow.

In addition, Bayesian neural networks \citep[e.g.][]{BISHOP1997, 2015arXiv150505424B} are not yet serviced by NS, meaning that computing neural model evidences is not yet possible. 
There are many high-quality open-source machine learning backends available, e.g. TensorFlow \citep{tensorflow2015-whitepaper}, PyTorch \citep{NEURIPS2019_9015}, JAX \citep{jax2018github}, Theano \citep{2016arXiv160502688T}, and CNTK \citep{10.1145/2939672.2945397}.
All of these machine learning backends have probabilistic programming packages which provide MCMC methods, e.g. NUTS-HMC \citep{2011arXiv1111.4246H}, however none yet have an implementation of NS.
If NS was available in any of these machine learning backends this would enable things like neural network model selection based on evidence, and Bayesian neural networks that are able to handle the multi-modal parameter posteriors that MCMC methods notoriously struggle with. 
Note, that even if any of the current NS implementations were able to be interfaced with a machine learning backend, they would still be too slow to be of practical use.

Given the above motivation for a high-performance NS implementation, we next logically must ask why the current implementations are slow.
Of the four NS implementations listed above, PolyChord and MultiNEST are written in Fortran with python bindings, DIAMONDS is written in C++, and dynesty is written purely in Python with NumPy and SciPy.
We do not consider DIAMONDS further in this paper.
Of the three remaining, PolyChord and dynesty have similar performance characteristics, and MultiNEST is slightly faster in lower dimensions becoming exponentially slower in higher dimensions.
However, all implementations are still too slow to meet anything except the offline-type inference applications mentioned above.
Note, it is possible to speed up these NS implementations via parallelising the likelihood evaluation, at the cost of multiplying computational resources.
The improvement is sub-linear, and it's unlikely that a typical user would have the computational resources to provide the required parallelisation.

There are two possible complimentary reasons these NS implementations are still slow despite being compiled.
The first is that perhaps the algorithms are inefficient with likelihood evaluations.
The number of likelihood evaluations required to get one effective sample is a good measure of performance for comparing NS implementations.
However, all of the NS implementations are based on innovative ideas to improving the efficiency of sampling, and it is unlikely that this is the dominant source of slowness.
The second reason is that the code is not optimised as much as it could be.
It is often difficult to well-optimise a large algorithm like NS.
In particular, it is typically difficult to achieve good memory locality, reduce memory fragmentation, and reduce the use of intermediate memory buffers.
While much of this is typically in the wheel-house of compilers, there is still much that a compiler cannot do without placing limitations on how the coder writes code.

These above considerations are the motivations for our high-performance NS implementation, which we call JAX-based Nested Sampling, or JAXNS, and is pronouced `jacksons'.
The high-performance of JAXNS follows from its implementation entirely in Google TensorFlow's Accelerated Linear Algebra\footnote{XLA: \url{https://www.TensorFlow.org/xla/architecture}} (XLA) primitives using JAX.
XLA is a maturing compiler designed by Google's compiler team to improve execution speed, improve memory usage, reduce reliance on custom kernels, and improve portability between accelerators of programs consisting largely of linear algebra.
XLA improves execution speed by compiling sub-graphs, fusing pipelined operations, constant folding, and apply a large barrage of expression substitutions.
In addition, XLA aggressively analyses memory allocation and performs intelligent scheduling. 
This results in less memory overhead, since intermediate memory buffers rarely need to be created, reduces memory fragmentation, and reduces cache misses via better locality.
Despite its name, XLA suppports much more than just linear algebra, such as conditional control flows, e.g. while loops and if-statements, and fancy indexing.

The most important optimisation in XLA is operation fusion.
For an example of how XLA might optimise a simple computation graph, consider \texttt{sum(x + y * z)}.
Without XLA, three operations would be launched: one for the multiplication, one for the addition and one for the summation.
However, with XLA the result is computed in a single operation, by fusing the addition, multiplication and summation into a single operation. 
This fused operation does not create intermediate values for \texttt{y*z} and \texttt{x+y*z}, rather it streams these intermediate values directly through the summations while leaving the input values entirely in registers. 
Since memory bandwidth is typically the scarcest resource on hardware accelerators, removing these intermediate memory operations is one of the best ways to improve performance.

The objects that XLA operates on are high-level optimisation (HLO) representations of computation graphs.
A HLO representation represents a computation in a hardware independent manner, and allows the compiler to segregate platform independent from platform dependent optimisations, as well as scheduling.
The XLA compiler performs two main steps which operate on HLO representations.
The first step is to perform accelerator independent optimisations, such as operation fusion, and sub-graph compilation.
The second step is accelerator specific optimisations, such as shape-specific array padding and sharding, and finally code generation.
Currently, there are three accelerator backends defined in XLA: CPU, GPU, and Google's TPU. 
Note, that XLA is currently the only way to write code for TPUs.

In order for XLA to perform these types of optimisations XLA requires the programs to be static-memory.
That is, the shapes and data types of arrays must be known at compile time.
This precludes algorithms that require growing arrays arbitrarily.
This aspect becomes particularly important in a dynamic algorithm such as NS, and is an important aspect in JAXNS's implementation design.

At its core JAX is simply an interface to the XLA compiler with some machinery to efficiently dispatch compiled XLA programs to distributed accelerator devices, as well as allow arbitrary order auto-differentiation.
JAX writes and reads almost exactly like NumPy and SciPy code, making it very user friendly for a large part of the scientific community, and as a result its user base is very quickly growing.

This paper is ordered in the following manner.
In Section~\ref{sec:review} we review the idea of NS and the basic components of the algorithm.
In Section~\ref{sec:implementation} we describe the static-memory implementation, and the particular details of the likelihood constrained sampler, as well as describe the API in Section~\ref{sec:api}.
In Section~\ref{sec:performance} we compare JAXNS to PolyChord, MultiNEST, and dynesty on some toy problems, and report the performance. 
Finally, we conclude with future plans and a hope for NS to become more common in Bayesian machine learning problems outside of astronomy.

\section{Nested sampling review}
\label{sec:review}

We'll now provide a summary of how NS computes Eq.~\ref{eq:1} and provides samples of the posterior distribution.
Note, that we'll use a notation of past NS literature, which can be confusing, and sometimes use the same symbol for variables and functions, letting the reader infer their meaning from the context.
The ingenuity of NS rests in reparametrising the domain of integration in terms of the enclosed prior mass inside of a likelihood contour level $L$, $X(L) = \int_{\mathcal{L}(x)>L} \, \mathrm{d} p(x)$.
By it's construction we have that the zero-likelihood contour encloses all prior mass, $X(0) = 1$, and the maximum likelihood contour encloses zero prior mass, $X(\max_x {\mathcal{L}(x)}) = 0$.
We have that $X(L)$ is a strictly monotonically decreasing function, therefore its inverse exists and is the likelihood as a function enclosed prior mass, $\mathcal{L}(X) = X^{-1}(X)$.
We term $L(X)$ the reparametrised likelihood function.
Since $X(L)$ is strictly monotonically decreasing, the reparametrised likelihood must be strictly monotonically increasing.
Under this reparametrisation, Eq.~\ref{eq:1} is given by,
\begin{align}
    Z =& \int_0^\infty X(L) \, \mathrm{d}L\label{eq:2}\\
    =& \int_0^1 \mathcal{L}(X)\,\mathrm{d}X.\label{eq:3}
\end{align}
This reparametrisation turns an integral over the, potentially high-dimension, prior space into a one-dimensional integral over enclosed prior mass, which seems to be a significant reduction in compute effort.
However, since we do not know $X(L)$ \textit{a priori} we must spend compute effort on estimating it.
\citet{2004AIPC..735..395S} made the observation that $X(L)$ could be estimated via a shrinkage process, which he called nested sampling and is encapsulated in this procedure,
\begin{enumerate}
    \item Set $t=0$, $X_0 = 1$, and $L_0=0$.
    \item Uniformly sample $n$ `live'-points from the likelihood-constrained prior, $P_t=\{x_i:i=1..n, x_i \overset{\rm i.i.d.}\sim p(x), \mathcal{L}(x_i) > L_t\}$.
    \item Set $X_{t+1} \overset{\rm i.i.d.}\sim X_t \mathrm{Beta}(n, 1)$, and $L_{t+1}=\min_{x\in P_t}\mathcal{L}(x)$.
    \item If a stopping criterion is not met, then set $t=t+1$, and go to step~(ii).
\end{enumerate}
The intuition behind this procedure is based on the order statistics of uniform random variables, and the fact that $L(X)$ is bounded and monotonically increasing.
Firstly, recall that if you have $n$ independent uniformly distributed numbers between zero and one, then the largest of these numbers will follow the $\mathrm{Beta}(n,1)$ distribution.
Now, in step (ii) the points in the set $P_t$ are uniformly sampled from the part of the prior enclosed by likelihood contour, $L_t$.
Since the likelihood is monotonically increasing as a function of enclosed prior mass, we can remove the point with the largest enclosed prior volume (lowest likelihood) and know that the enclosed prior mass will necessarily have shrunk.
The amount that the enclosed prior mass will have shrunk by is equal to the order statistic $\mathrm{Beta}(n,1)$, where $n$ is the number of points uniformly sampled.
Therefore, we can start at $X=1$, all enclosed prior mass, and walk in toward  $X=0$, zero enclosed prior mass, by successively sampling $n$ points within the current likelihood contour, shrinking $X$ by the correct amount, and replace the likelihood contour with the next smallest one.

\begin{figure}
    \centering
    \def\svgwidth{\columnwidth}
\begingroup%
  \makeatletter%
  \providecommand\color[2][]{%
    \errmessage{(Inkscape) Color is used for the text in Inkscape, but the package 'color.sty' is not loaded}%
    \renewcommand\color[2][]{}%
  }%
  \providecommand\transparent[1]{%
    \errmessage{(Inkscape) Transparency is used (non-zero) for the text in Inkscape, but the package 'transparent.sty' is not loaded}%
    \renewcommand\transparent[1]{}%
  }%
  \providecommand\rotatebox[2]{#2}%
  \newcommand*\fsize{\dimexpr\f@size pt\relax}%
  \newcommand*\lineheight[1]{\fontsize{\fsize}{#1\fsize}\selectfont}%
  \ifx\svgwidth\undefined%
    \setlength{\unitlength}{224.47651768bp}%
    \ifx\svgscale\undefined%
      \relax%
    \else%
      \setlength{\unitlength}{\unitlength * \real{\svgscale}}%
    \fi%
  \else%
    \setlength{\unitlength}{\svgwidth}%
  \fi%
  \global\let\svgwidth\undefined%
  \global\let\svgscale\undefined%
  \makeatother%
  \begin{picture}(1,0.71223867)%
    \lineheight{1}%
    \setlength\tabcolsep{0pt}%
    \put(0,0){\includegraphics[width=\unitlength,page=1]{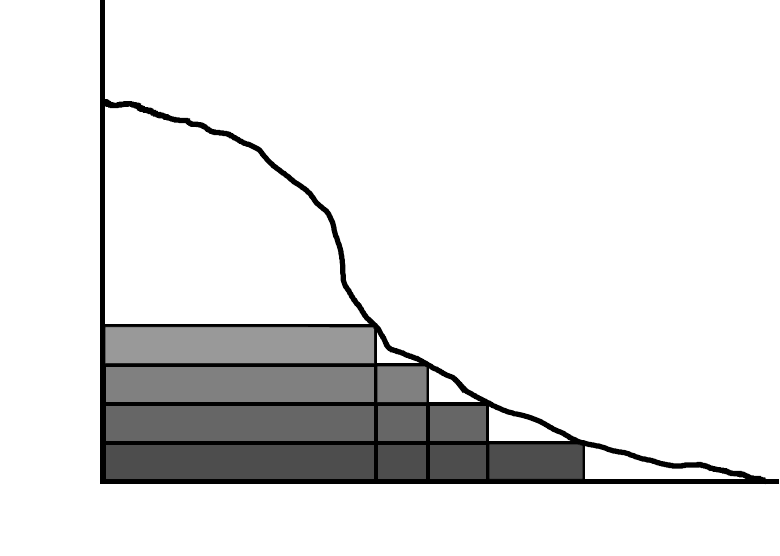}}%
    \put(0.32217746,0.30335815){\color[rgb]{0,0,0}\makebox(0,0)[lt]{\lineheight{1.25}\smash{\begin{tabular}[t]{l}$L$\end{tabular}}}}%
    \put(-0.00369784,0.39412339){\color[rgb]{0,0,0}\makebox(0,0)[lt]{\lineheight{1.25}\smash{\begin{tabular}[t]{l}$\mathcal{L}(X)$\end{tabular}}}}%
    \put(0.53224502,0.04401733){\color[rgb]{0,0,0}\makebox(0,0)[lt]{\lineheight{1.25}\smash{\begin{tabular}[t]{l}$X$\end{tabular}}}}%
    \put(0.19145205,0.00726517){\color[rgb]{0,0,0}\makebox(0,0)[lt]{\lineheight{1.25}\smash{\begin{tabular}[t]{l}$\{X:\mathcal{L}(X) > L\}$\end{tabular}}}}%
    \put(0,0){\includegraphics[width=\unitlength,page=2]{likelihood.pdf}}%
  \end{picture}%
\endgroup%

    \caption{A schematic depiction of Lebesgue integration performed by nested sampling, whose primary utility is to estimate the enclosed prior mass for a given likelihood contour level.}
    \label{fig:lebesgue_integral}
\end{figure}

This procedure can be interpreted as a probabilist's way of computing a Lebesgue integral, which can be seen by analogy with the Riemann integral.
The Riemann integral is the limiting process of partitioning the domain of a function into equal sub-intervals and summing the interval width times the height of the function in those intervals and taking the limit of infinitely many sub-intervals.
The Lebesgue integral, on the other hand, is the limiting process of partitioning the codomain of a function into contours of equal separation and summing the contour separation times the measure of the domain within that contour and taking the limit of infinitely many contours.
The shrinkage procedure achieves this by using order-statistics to estimate the measure of the domain within the contour.
This is depicted in Figure~\ref{fig:lebesgue_integral}.

This procedure will result in a sequence of $((X_t, L_t) : t=1..S)$ which can be used to numerically evaluate Eq.~\ref{eq:2} or Eq.~\ref{eq:3}, where $S$ is the number of samples required to terminate the algorithm.
In principle, this sequence is a random sequence due to the order statistic used to shrink the enclosed prior mass.
In practice, the expectation value is taken as the error incurred is typically small for typical values of $n$ \citep{2011MNRAS.414.1418K}.

The most difficult part of NS is the likelihood-constrained prior sampling.
The naive implementation of uniformly sampling $p(x)$ and rejecting all points that do not meet the constraint requires exponentially more likelihood calculations as the enclosed prior mass shrinks towards zero.
Most methods of likelihood-constrained prior sampling are based on bounding the set of live points and sampling within that bound, however not all are based on this approach.

The method used in MultiNEST is based on constructing multiple ellipsoids which bound the set of live points, and then rejection sampling within these ellipsoidal bounds taking into account ellipsoid overlap.
This method works well in many cases since many posterior distributions have ellipsoid structures.
However, the efficiency of this method suffers the curse of dimensionality, as the volume of prior space between the ellipsoidal bound and the true bound grows exponentially with dimension, and thus so too the rejection rate \citep{2015MNRAS.453.4384H}.
PolyChord improves over MultiNEST by doing away with the rejection sampling, instead incorporating a Markov chain approach with slice sampling generating proposals.
PolyChord also uses the live points to construct enclosed prior bounds, however it then uses these bounds to efficiently perform a sequence of one-dimensional slice sampling steps starting from one of the points in the live points until the user is satisfied that the proposals are decorrelated.
This Markov chain procedure has polynomial scaling with dimension.
PolyChord also utilises parallelisation to achieve a speed-up.

Other notable methods of likelihood-constrained prior sampling are constrained `Galilean' Hamiltonian Monte Carlo \citep{2011AIPC.1305..165B, 2013AIPC.1553..106F}, and diffusive nested sampling \citep{2009arXiv0912.2380B}.
Constrained Hamiltonian Monte Carlo sampling, like slice sampling, can be used to form a sequence of proposals inside the likelihood contour and therefore has polynomial scaling with dimension.
It requires computing the gradient of the likelihood several times per proposal, and reflecting off of the likelihood contours, which have the interpretation of infinite potential wells.
There is the added difficulty of choosing an appropriate step-size, which adds another tunable parameter.
Diffusive nested sampling is an interesting alternative class of nested sampling methods which rewrites the nested sampling procedure as the parallel evolution of particles.
Not all of these particles need to satisfy the likelihood constraint.
There seems to have been little development of diffusive nested sampling since its introduction.

\subsection{Marginalisation and parameter estimation}

The primary result of nested sampling is the sequence $\mathcal{S} = ((X_t, L_t) : t=1..S)$ where $S$ is the number of samples required to terminate the algorithm.
The evidence calculation uses $\mathcal{S}$ to approximate Eq.~\ref{eq:3}, however, consider the more general computation which is possible if the samples $x_t$ are also kept,
\begin{align}
    \mathbb{M}[f] \triangleq& \int_\mathcal{X} f(x) \mathcal{L}(x) \,\mathrm{d}p(x)\label{eq:4}\\
    \approx& \sum_t f(x_t) L_t (X_{t} - X_{t+1}).\label{eq:5}
\end{align}
Equation~\ref{eq:4} performs a marginalisation of a function $f:\mathcal{X} \to F$ over the unnormalised posterior distribution $\mathcal{L}(x) p(x) = Z p(y\mid x) p(x)$.
In this more general picture, the evidence is equivalent to $\mathbb{M}[1]$.
We can also use marginalisation to compute moments, e.g. the mean $Z^{-1} \mathbb{M}[x]$, the information gain $Z^{-1} \mathbb{M}[\log \mathcal{L}(x)] - \log Z$, and any other marginalisation of a bounded function.
From Eq.~\ref{eq:5} we see that the samples $\{ x_t : t=1..S\}$ are posterior samples with weights $\{w_t= Z^{-1} L_t (X_{t} - X_{t+1}) \}$.
Finally, we note that in Eq.~\ref{eq:5} we have assumed the simplest quadrature rule, however other more accurate rules, such as the trapezoid rule, can be used instead.

\section{Implementation}
\label{sec:implementation}

The design of JAXNS consists of two main components, both of which are dynamic in their equivalents found in other NS implementations and therefore need to be turned into static-memory components.
The first component is the main control-flow loop, and the second is the likelihood-constrained sampler.
The main loop consists of repeatedly checking a termination condition, and if it's not met, then acquire a new sample from the likelihood-constrained sampler.
The termination condition of JAXNS is the same as in PolyChord, i.e. the main loop terminates at step $t$ when the remaining evidence in the live points, $Z_{{\rm live}, t} \triangleq \langle L_{\rm live}\rangle X_t$ is less than a fraction, $f_{\rm term}$, of the current evidence estimate, $Z_t$.

The main loop passes a state between iterations, and this state contains the all the aggregate results of the algorithm.
Since the control must be static-memory, the state must consist of arrays whose shapes and types are known at compile time.
The state contents depend upon the mode which JAXNS is run in.
There are two modes of operation: 1) sample collection and marginalisation, and 2) marginalisation only.
In the first mode of operation JAXNS will retain the samples gathered and thus enable exploring the weighted posterior.
The first mode therefore requires specifying a maximum number of samples to collect so that arrays can be allocated at compile time.
In addition, JAXNS will perform all requested marginalisations incrementally as new samples arrive, i.e. marginalisation does not need to retain samples.
In the second mode of operation JAXNS will only perform marginalisation, and no samples will be retained.

The first mode is useful when the posterior is the desired product, or when an accurate calculation of the evidence uncertainty is desired.
Since the shrinkage of the enclosed prior mass at every iteration is a random variable, all marginalisations, Eq.~\ref{eq:5}, have an uncertainty component due to the trajectory of the shrinkage process.
In general, as the number of live points grows this trajectory uncertainty shrinks, however sometimes an accurate estimate of the evidence and its uncertainty is desired for model comparison.
This can be accomplished via stochastic simulation of shrinkage trajectories as suggested in \citet{skilling2006}.

One small caveat exists with sample collection.
When the information gain, $\mathcal{I}$, is very large, i.e. the Kullback-Leibler divergence from the prior to the posterior is large, the number of samples required to reach termination can be very large.
The typical number of samples required to satisfactorily shrink the enclosed prior mass to zero is $n \mathcal{I}$, where $n$ is the number of live points \citep{2004AIPC..735..395S}.
It is not uncommon for $\mathcal{I} \approx 1000$ when there are many active constraints in a model.
Therefore, with large $n$ it is possible that the user does not (or cannot) allocate enough room for all the samples.
All but a small fraction of these samples will have negligible contribution to the evidence, and therefore one possible solution is reservoir sampling \citep[e.g.][]{chao1982} to perform online resampling and maintain a bounded list of samples.
This is part of the future roadmap of JAXNS.

To perform marginalisation, including evidence calculation, we apply the same online approach that PolyChord uses to compute the evidence, modified to account for arbitrary marginalisations.
Namely, we replace $L_t$ with $L_t f(x_t)$ in Appendix B of \citet{2015MNRAS.453.4384H}.
While the trajectory uncertainty has an effect on marginalisations as well, typically this uncertainty is less than the uncertainty of contour sampling and it can be neglected \citep{2011MNRAS.414.1418K}.

The second component of JAXNS is the likelihood-constrained prior sampling.
JAXNS implements a modified version of MultiNEST's bounding ellipsoidal construction to bound the live points, and then performs a modified version of PolyChord's slice sampling to propose new points within that bound.
We will assume the reader familiar with MultiNEST's and PolyChord's implementation details, and we will focus on the differences with JAXNS.

The bounding ellipsoids are calculated following a heirarchical process of splitting a subset of live points into two clusters until there is no gain in splitting.
This process is originally dynamic in nature, as the number of splittings is not known before hand.
In JAXNS the maximum number of clusters must be designated before hand.
Let $P$ be the set of all live points, $E(Q)$ be the bounding ellipsoid of a subset of points $Q \subset P$, and $V_E(Q)$ the volume of the bounding ellipsoid of $Q$.
The procedure of calculating the bounding ellipsoids is,
\begin{enumerate}
    \item Set $Q_0=P$, and $E_0=E(P)$.
    \item Assign $Q_k$ into $Q_{k,0}$ and $Q_{k,1}$ according to the modified k-means metric of \citet{2008MNRAS.384..449F} (described below).
    \item If maximum splittings is not yet reached and $V_E(Q_{k,1}) + V_E(Q_{k,2}) > V_E(Q_k)$ (ellipsoids intersect) then accept this assignment and apply steps (ii) and (iii) to each subset (letting the subset subscript become $k$).
\end{enumerate}
In MultiNEST the procedure of assigning points into subsets happens at once for all points.
We found that this leads to ill behaviour in some situations.
In JAXNS we instead start with vanilla 2-means clustering and then apply their reassignment metric greedily one element at a time until convergence (typically only a few iterations are needed).
The most important aspect in making this a static-memory algorithm is in the recursive aspect of the computation.
In order to make efficient use of JAX's control-flow we deterministically encode the order of subset splittings carried out by the recursive application of splitting.
This is done by letting each slot in the subset subscripts correspond to a bit, which enocdes a unique index of the unrolled splitting computation.
A while loop can then conditionally perform each splitting by rolling through the indices $0..2^{(\mathrm{depth}-1)}$, where $\texttt{depth}$ controls the maximum number of hierarchical clusters.
The splittings that did not occur are treated as zero-volume ellipsoids.

We use the bounding ellipsoids to guide slice sampling, however our implementation deviates from PolyChord in several ways.
Since slice sampling \citep{2000physics...9028N} is a Markov chain method, it is naturally a static-memory algorithm, and our modifications are only chosen to improve performance.
PolyChord performs clustering and attempts to locate and track posterior modes semi-independently.
It uses \textit{phantom points}\footnote{Including a variable number of phantom points would break the static-memory requirements.}, correlated intermediate samples of slice sampling, to help characterise the shapes of the posterior modes, and then applies an affine whitening transformation to points within a posterior mode before slice sampling. 
If the posterior mode is accurately located and characterised, then this means that the initial bracketing interval will be accurate, and slice sampling inside that mode will be efficient.
However, performing the whitening process requires computing the Cholesky and performing a triangular-matrix solve on every iteration.
Furthermore, if a slice tunnels to a different posterior mode, then the whitening needs to happen again, invoking further computation.

\begin{figure}
    \centering
    \includegraphics[width=\columnwidth]{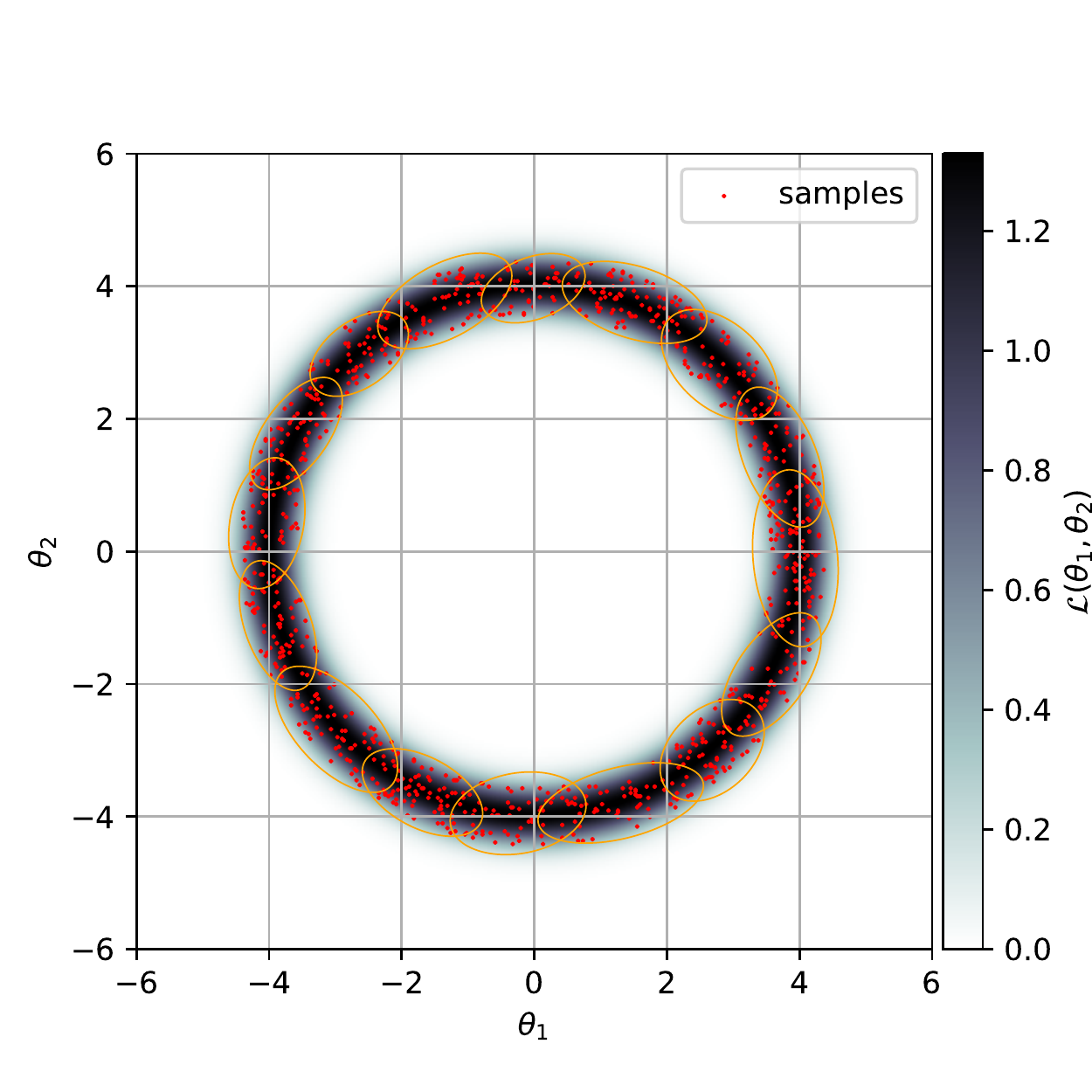}
    \caption{Ellipsoidal clustering of a two-dimensional Gaussian shell likelihood with \texttt{depth=5} (maximum number of clusters is 16).}
    \label{fig:gaussian_shells}
\end{figure}

Whitening also only makes sense when the posterior mode is a convex region, i.e. a straight line between any two points in the region is contained in the region. 
Otherwise the estimated bracketing interval, which is based on the covariance matrix of the points, can make little sense.
A simple example is a circular Gaussian shell likelihood with uniform prior \citep[e.g.][]{2009MNRAS.398.1601F}, shown in Figure~\ref{fig:gaussian_shells}.
An intersection of ellipsoids easily bounds the posterior structure, and the posterior mode can be identified because the ellipsoids are intersecting.
Due to the large hole in the centre of the points, performing whitening in such a situation would result in an bracketing interval that is much too large, and many shrinkage steps would need to be taken, decreasing efficiency. 
In addition, many of the samples would fall in the hole region.
Therefore, our three main differences with PolyChord are that we do not whitening, do not attempt to track posterior modes, and we do not keep phantom points.

Instead, our slice sampling is summarised as follows.
Let $\{E_k : k=1..K\}$ be the ellipsoidal clustering of the live points $P$, and let $\# : \mathcal{X} \to \mathbb{N}$ assign an ellipsoid to a point in the prior space, then our slice sampling procedure is encapsulated by,
\begin{enumerate}
    \item Uniformly select a starting point, $x_0 \in P$.
    \item Uniformly sample a direction $\hat{n}_i \in \mathbb{S}^{D-1}$.
    \item Use $x_i$, $\hat{n}_i$, and $E_{\#(x_i)}$ in the modified step-out and shrinkage procedure (described below) to find a new point, $x_{i+1}$.
    \item If a user-supplied termination condition is not met, then go to step (ii), otherwise return $x_{i+1}$.
\end{enumerate}

Since we do not track posterior modes, we cannot use them to pick the initial point to slice sample from.
This is different from both MultiNEST and PolyChord which select the initial cluster/point to sample from based on the estimate of the posterior mode volume.
In JAXNS the initial point is selected randomly, and we accept the random-walk behaviour of posterior mode volumes discussed in \citep{2015MNRAS.453.4384H}. 
Since we do not attempt to estimate the volume of individual posterior modes, this is not a practical issue for JAXNS.
In addition, this random-walk behaviour is partially offset by the ability of live points to tunnel between nearby ellipsoids.
We address pan-NS implementation problem of posterior mode evaporation by increasing the number of live points.
Note, that the validity of slice sampling procedure as outlined in \citet{2000physics...9028N} does not depend on choosing the initial point based on the volume of individual modes. 

In contrast with PolyChord, during the step-out and shrinkage procedure of each one-dimensional slice sampling, we do not perform whitening.
Instead, we calculate the points of intersection of the slice ray and the bounding ellipsoid and use this to set the initial window size of the step-out procedure.
The shrinkage procedure is the same as described in \citet{2000physics...9028N}.
We also compute the intersection of the ray with the prior domain boundary, so that a slice may never probe outside of the prior domain.

After accepting a slice sampling proposal we accordingly shrink the bounding ellipsoids and check a condition of whether or not to reconstruct the bounding ellipsoids.
The condition used in MultiNEST is based on the deviation of the volume of the bounding intersecting ellipsoids from the current estimate of the enclosed prior mass.
We found this condition to be too sensitive to the robustness of the ellipsoidal clustering method, and sometimes lead to repeated reclusterings.
The condition used in PolyChord is a user-defined number of samples, usually in multiples of the number of live points, $n$.
This has the interpretation that after $n$ samples the enclosed prior mass will have approximately shrunk by $e^{-1}$, and this might be a good time to readjust the clustering.
However, this static condition is not sensitive to temporary periods of very poor clustering, leading to lowered efficiency.
In JAXNS, the condition for recalculating the bounding ellipsoids is when the instantaneous likelihood evaluations per proposal is $2 D$ greater than the exponential moving average over the last $n$ samples, where $D$ is the dimension of the prior.
This condition means that we recalculate the ellipsoids when the slice sampler had to perform one extra step-out and shrinkage step in each dimension of the slice-sampler (which is what we term a drop in efficiency).
The exponential moving average gets reset after every recalculation of the ellipsoids.

\subsection{Prior domains}

As is common with all NS implementations, JAXNS prefers the user to parametrise the prior as a transformation from the uniform prior unit cube \citep[e.g.][]{2015MNRAS.453.4384H}, as this means that the prior domain is dimensionless, sometimes referred to as having units of probability. 
However, this parametrisation is not always possible.
In particular, some continuous random variables have troublesome transforms from the unit uniform distribution, e.g. the gamma distribution with non-integer shape parameter.
In addition, discrete random variables have no such transform.
JAXNS handles this by having separate prior domain subspaces for priors lacking a transformation from the unit cube, and letting the slice sampler handle these subspaces accordingly. 

One common feature missing from most NS implementations are discrete prior domains.
It is made clear in \citet{2004AIPC..735..395S} that nested sampling is entirely valid for discrete priors, as it only depends on the fact that the likelihood is monotonically increasing with respect to enclosed prior mass.
In addition, slice sampling is valid for sampling from discrete distributions \citep[e.g.][]{walker2007, maria2011}. 
At the time of publication, JAXNS does not have support for discrete variables, however it is in the future plans.

\subsection{JAXNS API and parameters}
\label{sec:api}
We provide a convenient probabilistic programming API with JAXNS that enables easily defining models.
Prior distributions may be hierarchical, i.e. they may depend on parameters which are also random variables.
The user must also specify the log-likelihood function as a function of the prior variables.
In addition, the user must supply any functions that should be marginalised, with the exception of the evidence and information gain, which are always calculated.
Many examples are available via the public repository of JAXNS\footnote{\url{https://www.github.com/joshuaalbert/jaxns}}.

The user must provide a number of parameters that define the NS run.
The first is \texttt{num\_live\_points}.
The number of live points should be based on the users expectation about the number of posterior modes, and the number of dimensions.
A good rule of thumb is \texttt{num\_live\_points=50*(\# posterior modes)*(D+1)}, where the prefactor of 50 is a rough figure and can be adjusted, and $D$ is the dimension of the prior. 
The second is \texttt{collect\_samples}, which is a Boolean that determines if samples should be retained, or if only marginalisation should be performed.
Performing only marginalisation can be faster in some situations, however in some situations, collecting the samples and computing the marginalisation after can be faster.
The third is \texttt{max\_samples}, an integer that defines the maximum number of iterations and samples that can be taken by NS. 
This should be large enough, and a good number is information gain times the number of live points times a few.
The fourth parameter is \texttt{depth}, an integer which determines the maximum number of ellipsoids to construct via hierarchical splitting.
This should be based on the user's expectation of the number of posterior modes
The fifth parameter is \texttt{num\_slices}, an integer which determines how many slice sampling proposals to perform, as a multiple of $D$, before accepting the proposal. 
Finally, the termination condition \texttt{termination\_frac}, corresponding to $f_{\rm term}$, must be chosen. 
A robust nominal setting for the latter three parameters is \texttt{depth}=3, \texttt{num\_slices}=5, and \texttt{termination\_frac=0.001}.

\section{Performance comparison}
\label{sec:performance}

We compare the performance of JAXNS to MultiNEST, PolyChord, and dynesty in two categories: the accuracy of the evidence calculations, and the wall-time required to compute the evidence.
We are interested in these metrics as a function of the dimension of the problem and the complexity of the posterior space.
We use two simple models to test these two attributes.
The first model is a highly-correlated multivariate Gaussian likielihood, with conjugate prior,
\begin{align}
    \boldsymbol{\theta} \sim& \mathcal{N}(\mathbf{0}, \mathbf{I})\\
    \mathcal{L}(\boldsymbol{\theta} ) =& \mathcal{N}(2 \mathbf{1}, \boldsymbol{\Sigma})\\
    \boldsymbol{\Sigma}_{ij} =& 
    \left\{\begin{matrix}
    1 & i=j\\
    0.95 & i\ne j
    \end{matrix}\right.,
\end{align}
where $\mathbf{1}$ is a vector of ones, and $\mathbf{I}$ is the identity matrix.
The evidence of this model is,
\begin{align}
    Z = \mathcal{N}(2\mathbf{1} \mid \mathbf{0}, \boldsymbol{\Sigma} + \mathbf{I}),
\end{align}
where we use the notation $\mathcal{N}(\cdot \mid \mu, \Sigma)$ to mean the probability density function with mean $\mu$ and covariance $\Sigma$.

The second model's likelihood is a mixture of $2 K$ highly-correlated multivariate Gaussians, with a wide multivariate Gaussian prior,
\begin{align}
    \boldsymbol{\theta} \sim& \mathcal{N}(\mathbf{0}, K^2\mathbf{I})\\
    \mathcal{L}(\boldsymbol{\theta} ) =&  (2 K)^{-1} \sum_k^K \mathcal{N}(\boldsymbol{\mu}^k, \boldsymbol{\Sigma}^-) + \mathcal{N}(\boldsymbol{\mu}^k, \boldsymbol{\Sigma}^+)\\
    \boldsymbol{\mu}^k_i=& 1 + k\\
    \boldsymbol{\Sigma}^\pm_{ij} =& 
    \left\{\begin{matrix}
    0.5 \pm 0.49 & i=j=0\\
    0.5 \mp 0.49 & i=j\ne 0\\
    0&i\ne j
    \end{matrix}\right..
\end{align}
The evidence of this model is given by,
\begin{align}
    Z =& (2 K)^{-1} \sum_k^K \mathcal{N}(\boldsymbol{\mu}^k \mid \mathbf{0}, K^2\mathbf{I} + \boldsymbol{\Sigma}^-) + \mathcal{N}(\boldsymbol{\mu}^k \mid \mathbf{0}, K^2\mathbf{I} + \boldsymbol{\Sigma}^+).
\end{align}
\begin{figure}
    \centering
    \includegraphics[width=\columnwidth]{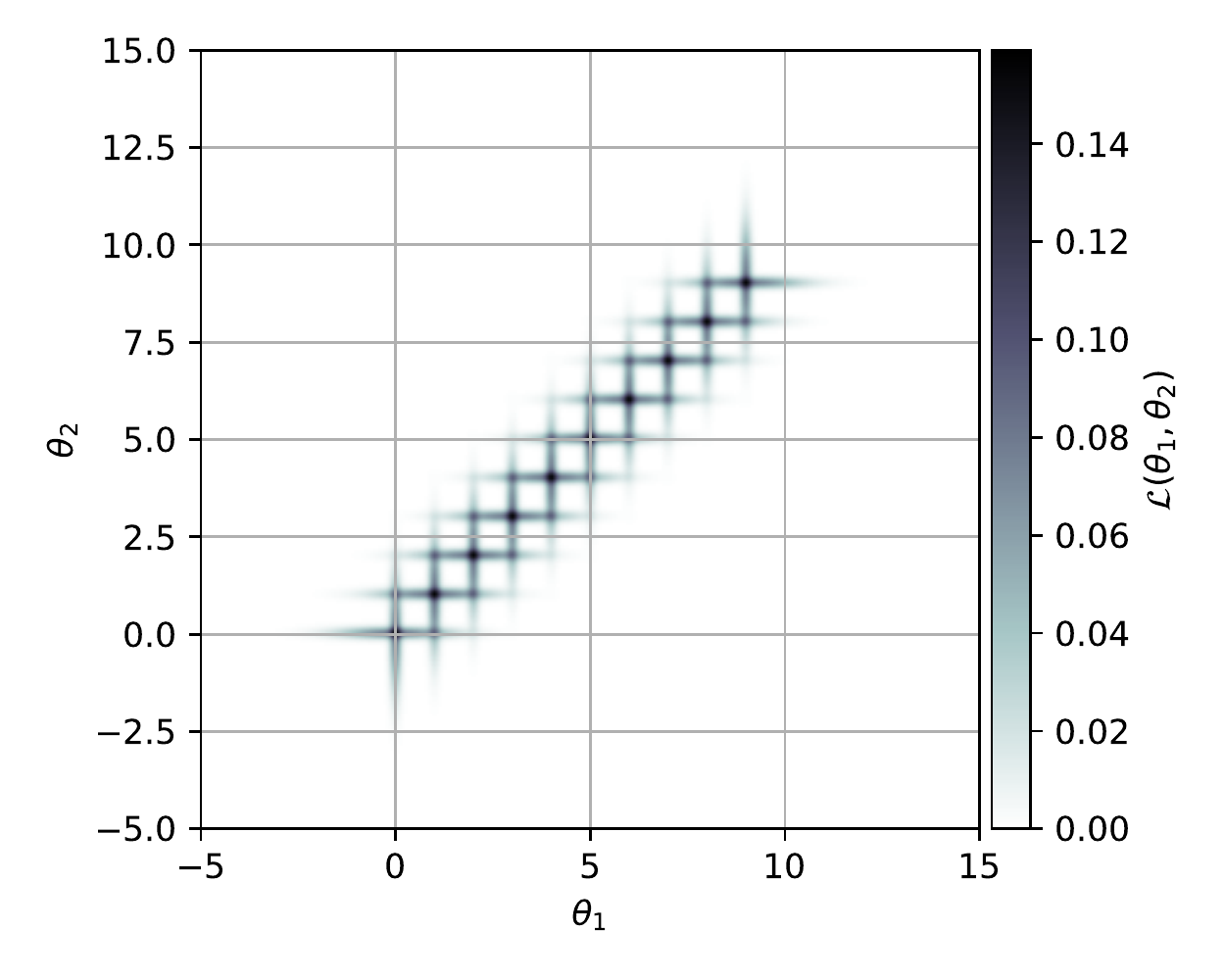}
    \caption{Second model's likelihood in two-dimensions, with $K=10$.}
    \label{fig:second_model_likelihood}
\end{figure}
The likelihood of the second model is an overlapping group of multi-dimensional cross-hatches, and is depicted for two-dimensions in Figure~\ref{fig:second_model_likelihood}.

Since JAXNS is a static-memory algorithm it is not possible to use queues and thread pools to parallelise the likelihood evaluations, however there is a static-memory variant that may be incorporated in the future.
Therefore, for this comparison we run all the NS implementations in their single thread-pool variants for equal comparison.
Furthermore, since we understand how distributing likelihood evaluation over slave processes speeds up NS we can reason with just their single-thread variant performances.
For posterity, we have that with $n_{\rm proc}$ slave processes evaluating likelihoods the average gain in useful likelihood evaluations goes as, $\langle n_{\rm gain} \rangle = n_{\rm live} \log(1+n_{\rm proc}/n_{\rm live})$ \citep{2015MNRAS.453.4384H}.

We attempt to set the NS implementations to have similar quality settings.
We use the dynesty slice sampler \texttt{sampler=`slice'}, and MultiNEST importance NS \citep{2019OJAp....2E..10F}.
We control the uniformity of the samples for JAXNS, PolyChord, and dynesty by taking $5D$ one-dimensional slice proposals per sample. 
For MultiNEST the uniformity is guaranteed so long as the the enclosed prior subdomain is properly constrained, which is controlled by \texttt{sampling\_efficiency=0.3}, a good setting for computing evidences.
For JAXNS and PolyChord we use the same $f_{\rm term}=0.001$ termination condition, for dynesty we use \texttt{dlogz=0.01}, and for MultiNEST we use \texttt{evidence\_tolerance=0.5}. 
Regarding clustering, for PolyChord we turn clustering on for all experiments via \texttt{do\_clustering=True}, for MultiNEST set \texttt{max\_modes=100}, and for JAXNS set \texttt{depth=3} in the first model and \texttt{depth=7} in the second model.
For dynesty we use \texttt{bound=`single'} for the first model, and \texttt{bound=`multi'} for the second model.

\begin{figure}
    \centering
    \includegraphics[width=\columnwidth]{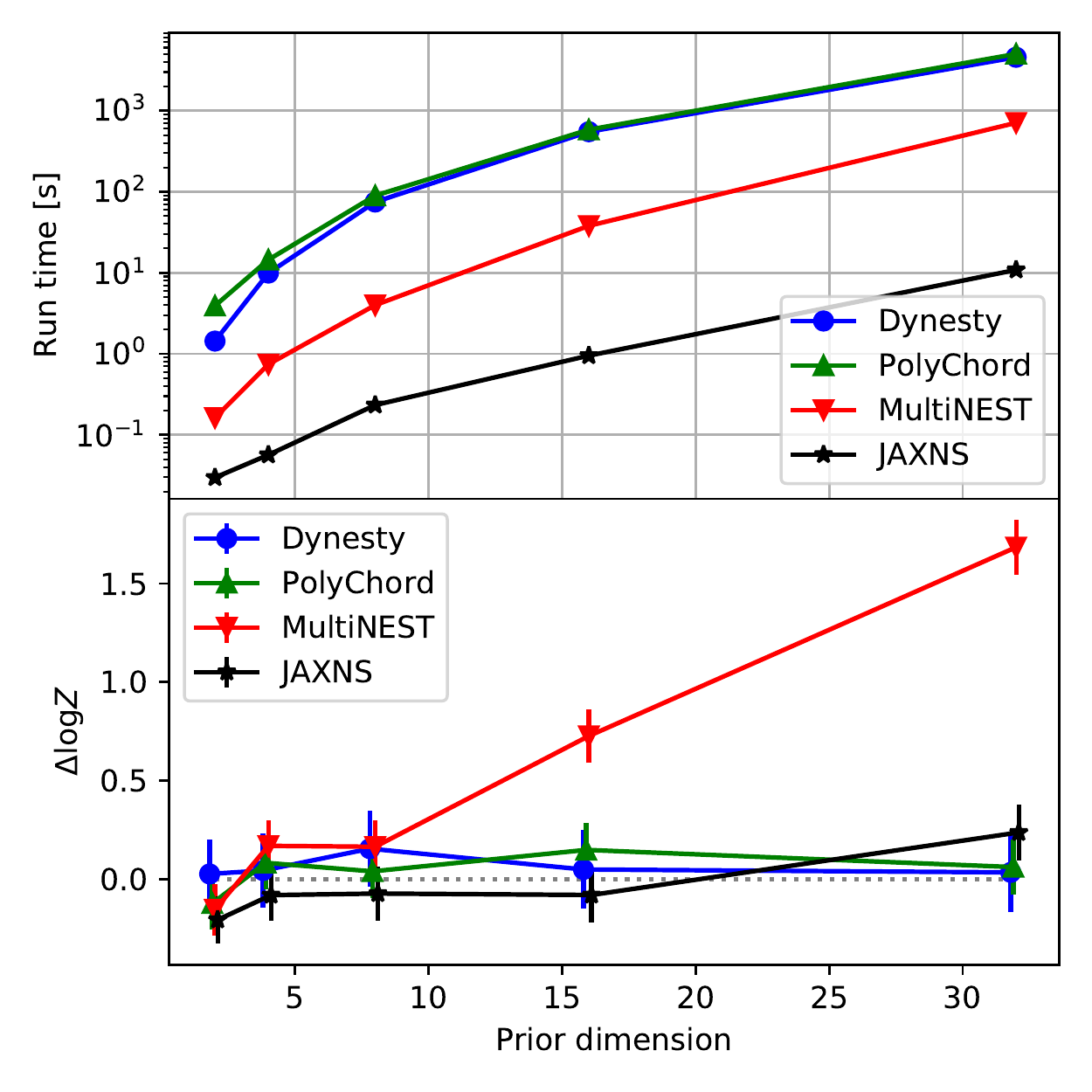}
    \caption{The run time (top) and evidence accuracy (bottom) of the first model for the different NS implementations.
    The evidence accuracy is plotted as the different between the mean computed evidence and the true evidence.}
    \label{fig:performance_first}
\end{figure}

\begin{figure}
    \centering
    \includegraphics[width=\columnwidth]{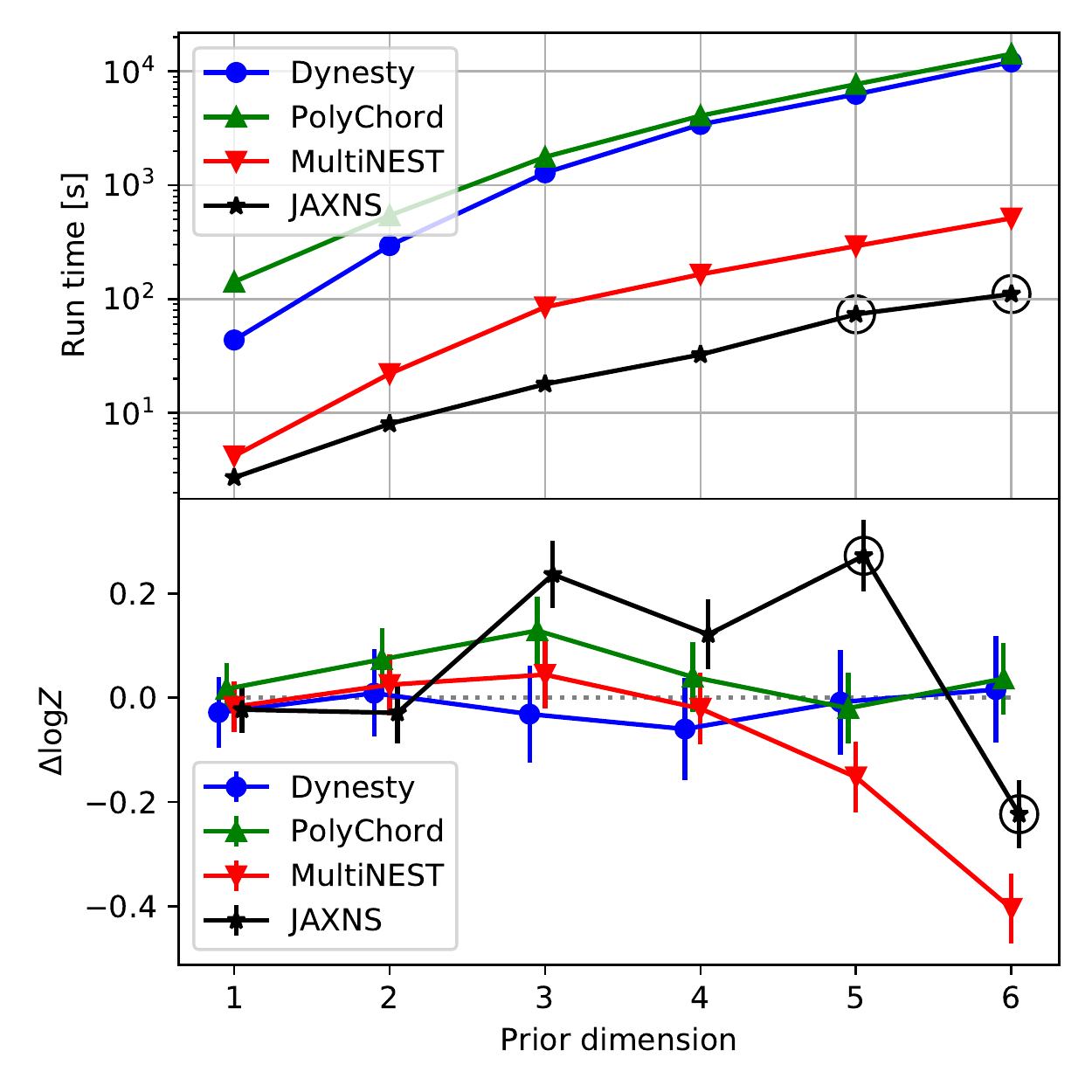}
    \caption{The run time (top) and evidence accuracy (bottom) of the second model for the different NS implementations.
    The evidence accuracy is plotted as the different between the mean computed evidence and the true evidence. 
    The black circles at $D=5$ and $D=6$ indicate JAXNS runs with \texttt{depth=8} and \texttt{depth=9}, respectively.}
    \label{fig:performance_second}
\end{figure}

For the first model we run $n=50 D$ live points, and for the second model we run with $n=50 D K$ live points, where $D$ is the dimension of the prior domain and $K$ is the number of posterior modes (mixture components).
In these experiments we chose $K=10$.
The evidence accuracy and run time of the first model is shown in Figure~\ref{fig:performance_first}, and for the second model in Figure~\ref{fig:performance_second}.

We observe that in the first model JAXNS is between one and two orders of magnitude faster than MultiNEST and two to three orders of magnitude faster than PolyChord and dynesty.
Furthermore, as dimension increases the difference with MultiNEST increases, since slice sampling has polynomial scaling and MultiNEST has exponential scaling.
Note, that for $D=64$ dynesty and PolyChord took longer than a day to compute, so we stopped the experiment at $D=32$.
In the second model, where the posterior is significantly more complicated, JAXNS is approximately one order of magnitude faster than MultiNEST, and two orders of magnitude faster than PolyChord and dynesty.

In the evidence plot of the first model we observe that dynesty, PolyChord, and JAXNS have similar evidence accuracy, and have error consistent with their deviation from the true evidence.
However, we find that MultiNEST's evidence calculations seem to deviate as a function of prior dimension.
It's unclear where this error in MultiNEST's evidence calculation is coming from.
In the second model, all the models are centred around the true evidence, however, for $D = 5$ and $D=6$, JAXNS's evidence error was significantly larger than the estimated uncertainty when run with \texttt{depth=7} and we had to increase it to \texttt{depth=8} and \texttt{depth=9}, respectively, for these runs.
The large evidence error follows from non-uniform sampling, e.g. incorrect live point bounds, and only becomes present in highly complicated posteriors.
When we increased the \texttt{depth} (black circles) the run times increased slightly and the evidences became more accurate.
Setting \texttt{depth} \textit{a prior} is a side-effect of the static-memory requirement, and this problem illustrates one limitation due to that. 

\section{Discussion}
\label{sec:discussion}

The heavily increased performance of JAXNS can be attributed to its static-memory implementation in XLA primitives.
This level of speed up due to XLA is also exhibited in other realms of machine learning.
For example, in the Machine Learning Performance v0.7 benchmarks of July 2020 the world saw large deep learning models train from scratch in 30~seconds, where the same models on the top hardware in 2015 took 3 weeks.
This constitutes a five order of magnitude improvement, apparently significantly breaking Moore's law.
However, this speed up was mostly attributed to the software improvements since 2015, i.e. XLA from Google and similar related products from nVidia and others.
Therefore, it is not unexpected that JAXNS should be much faster than NS implementations written in older frameworks.

There is a growing sphere of probabilistic programming frameworks among the various machine learning backends, e.g. TensorFlow Probability for TensorFlow, PyRo for PyTorch, pymc3 for theano, NumPyRo for JAX, and so on.
Prior to JAXNS, none of these backends had a NS implementation, and thus a way of calculating evidences.
We believe there is a legion of potential applications of NS in machine learning.
For example, the author would be excited to see researchers use evidence calculations as reward signals in reinforcement learning.
In addition, sub-blocks of neural network parameters could be treated in a Bayesian way and their complicated posteriors accurately inferred with NS.

Another exciting realm of NS application is in Gaussian process regression with hyper parameter marginalisation \citep{2020MNRAS.497.4672A, 2020arXiv201016344S}.
Currently, the popular tool for Gaussian process inference is evidence (or variational bounds of evidence) maximisation \citep[e.g.][]{2016arXiv161008733M}, thus working with point-wise estimates of hyper parameters.
The author would be happy to see NS become the default for Gaussian process inference, in particular, we see it as the important next step in Bayesian optimisation \citep[e.g.][]{2017arXiv171103845K}.

\subsection{Future plans}
\label{sec:future}

At the moment JAXNS is limited by the static-memory requirement of placing a cap on the number of collected samples.
This means that the algorithm will terminate when this number of samples are collected, potentially before all the evidence is calculated.
However, since most of the collected samples have very small weights, a significant improvement would be to keep only a selection of the samples that have significant weights.
We plan on introducing a static-memory variant of reservoir sampling \citep[e.g.][]{chao1982} that enables online resampling of the new nested samples and maintaining an equally weighted reservoir of samples that preserves the statistics of the entire sample population.
This improvement would not effect marginalisation.

Nested sampling is a powerful inference method that works for both continuous and discrete random variables, as shown originally in \citet{2004AIPC..735..395S}.
This makes it a potentially valuable tool for various physics models, and plausible reasoning.
We have planned an extension of JAXNS to enable discrete priors.

Even if parallelised likelihood evaluation were used by dynesty, PolyChord, or MultiNEST, they we still need between 100 and 1000 slave processes to achieve the speed of JAXNS with sequential likelihood evaluation.
While the use of a likelihood-evaluation queue would break the static-memory requirement, we have a planned static-memory variant which will increase the efficiency of likelihood evaluations.
Thus, we expect that once this is implemented JAXNS should become scalably faster with a controllable computation-efficiency gain trade-off.

\section{Conclusions}
\label{sec:conclusions}

We have reformulated nested sampling as a static-memory algorithm and implemented it with XLA primitives using JAX, a high-performance backend for accelerator-independent machine learning.
In comparison with the nested sampling packages MultiNEST, PolyChord, and dynesty run in their sequential likelihood evaluation variants, our implementation is two to three orders of magnitude faster, while maintaining the same accuracy of evidence calculations.
We have released our implementation in an open-source package called JAXNS (pronounced `jacksons').
Since it is written entirely in JAX, it provides a powerful tool to the machine learning community, and the author hopes it enables new evidence-based explorations in many fields.

\section*{Acknowledgements}

JA would like to thank his colleague, and friend, Martijn Oei for helpful discussions.

\section*{Data Availability}

All code and data required to reproduce the results of this study are provided in a git repository \url{https://www.github.com/joshuaalbert/jaxns_paper}.



\bibliographystyle{mnras}
\bibliography{cite, non_ads_cite}



\bsp	
\label{lastpage}
\end{document}